\documentclass[twocolumn,prl,floatfix,citeautoscript,nofootinbib,superscriptaddress]{revtex4-2}
\usepackage{amsbsy}
\usepackage{latexsym,epsfig,graphicx}
\usepackage{dcolumn}
\usepackage{graphicx}
\usepackage{subfigure}
\usepackage{comment}
\usepackage{color}
\usepackage{bm}
\usepackage{mathrsfs}
\usepackage{amsfonts}
\usepackage{amsmath}
\usepackage{color}
\usepackage{amssymb}
\usepackage{xspace}
\usepackage{epstopdf}
\usepackage{tabularx}
\usepackage{longtable}
\usepackage[colorlinks=true, letterpaper=true, pdfstartview=FitV, linkcolor=red, citecolor=blue, urlcolor=blue]{hyperref}
\usepackage[normalem]{ulem}

\pdfoutput=1

\begin{document}

\title{Distributed quantum computing with photons and atomic memories}
\author{Eun Oh}
\affiliation{Arcadia Drive, Ellicott City, Maryland 21042, USA}
%\affiliation{JTEC Consulting, Decatur, Georgia 30030, USA} % For arXives, use this.  For PRL, use LPS.
%\affiliation{Laboratory for Physical Sciences, College Park, Maryland 20740, USA}
\author{Xuanying Lai}
\affiliation{Department of Physics, The University of Texas at Dallas, Richardson, Texas
75080, USA}
\author{Jianming Wen}
\affiliation{Department of Physics, Kennesaw State University, Marietta, Georgia 30060, USA}
\author{Shengwang Du}
\email{dusw@utdallas.edu}
\affiliation{Department of Physics, The University of Texas at Dallas, Richardson, Texas
75080, USA}

%\date{\today }

\begin{abstract}
The promise of universal quantum computing requires scalable single- and inter-qubit control interactions. Currently, three of the leading candidate platforms for quantum computing are based on superconducting circuits, trapped ions, and neutral atom arrays. However, these systems have strong interaction with environmental and control noises that introduce decoherence of qubit states and gate operations. Alternatively, photons are well decoupled from the environment, and have advantages of speed and timing for distributed quantum computing. Photonic systems have already demonstrated capability for solving specific intractable problems like Boson sampling, but face challenges for practically scalable universal quantum computing solutions because it is extremely difficult for a single photon to ``talk'' to another deterministically. Here, we propose a universal distributed quantum computing scheme based on photons and atomic-ensemble-based quantum memories. Taking the established photonic advantages, we mediate two-qubit nonlinear interaction by converting photonic qubits into quantum memory states and employing Rydberg blockade for controlled gate operation. We further demonstrate spatial and temporal scalability of this scheme. Our results show photon-atom network hybrid approach can be an alternative solution to universal quantum computing.
\end{abstract}

\maketitle

% \emph{{\color{blue}Introduction}.---}
\emph{{\color{blue}Introduction}.---}Different from bits (0 and 1) in a classical digital computer, a quantum bit (\textit{i.e.} qubit) is generally a superposition of two discrete states $|0\rangle$ and $|1\rangle$ and multiple qubits can be quantum mechanically entangled.  Analogous to digital gates in a classical computer, a universal quantum computer also requires a set of basic quantum gates to operate its qubits \cite{QI-Barnett}. These universal gates can be rotation operators, phase shift and controlled-NOT (CNOT) gates \cite{Williams2011}. Currently, there are three leading candidates for quantum computers platforms: superconducting circuits \cite{npjQI-Matthias}, trapped ions \cite{IonQC-Sage}, and neutral atom arrays \cite{PhysicsToday-Saffman, Saffman_2016}. Even though there are on-going efforts to address various challenges, all these systems have strong interactions with environmental and control noises that introduce decoherence and limited lifetime for quantum computation \cite{RevModPhys.75.715, RevModPhys.76.1267}.  

On the contrary, photons are well decoupled from the background, travel at the highest speed in the universe and can be precisely controlled in picosecond time resolution routinely in lab. Recently, photonic systems have demonstrated power in solving intractable problems like Boson sampling \cite{ScienceLu2020}, but face challenges for practically scalable universal quantum computing solutions because it is extremely difficult for a single photon to control another deterministically. Though manipulating photonic single qubit is straightforward with linear optics including wave plates, mirrors, and beam splitters \cite{Barz_2015, Saleh1991}, the path towards universal quantum computer faces a great challenge due to lack of efficient optical nonlinearity at a single-photon level. The widely used scheme with linear optics, making use of probabilistic measurement induced effective ``nonlinearity", is practically not efficient for large scale implementation because it requires enormous amount of ancilla photons and the computational time scales exponentially with the number of gates \cite{Nature.409.46, RevModPhys.79.135}. 

Atomic ensemble Rydberg state mediated nonlinearity has been proposed and demonstrated for realizing photon-photon interaction gates, which requires conversion between photonic states and collective Rydberg polariton states \cite{PhysRevLett.87.037901, PhysRevLett.112.040501, NatPhys.15.124}. However, such a quantum memory (QM) with Rydberg polaritons has very low storage-retrieval efficiency ($<$ 10\% reported) \cite{NatPhys.15.124, PhysRevLett.110.103001} which limits its practical applications. In this Letter, we propose a universal quantum computing scheme based on photonic polarizations and efficient atomic-ensemble ground-state QMs. To introduce nonlinear interaction between two qubits, we convert the photonic qubit states into atomic-ensemble-based QM states and implement two-qubit controlled-phase (CP) gate with Rydberg blockade effect. %The readout photons are then fed to next layer quantum circuit. %This scheme is different from the recent demonstration of the non-degenerate photon-photon controlled phase flip gate based on Rydberg interaction induced modification on the linear electric susceptibility in which a control (or target) photon cannot interact with another control (or target) photon and thus its scalability is limited \cite{NatPhys.15.124}.

%Different from the widely implemented single neutral atom based quantum computing platforms, here the key elements are efficient atomic ensemble based quantum memories. 

\begin{figure}[t]
\includegraphics[width=1.0\linewidth]{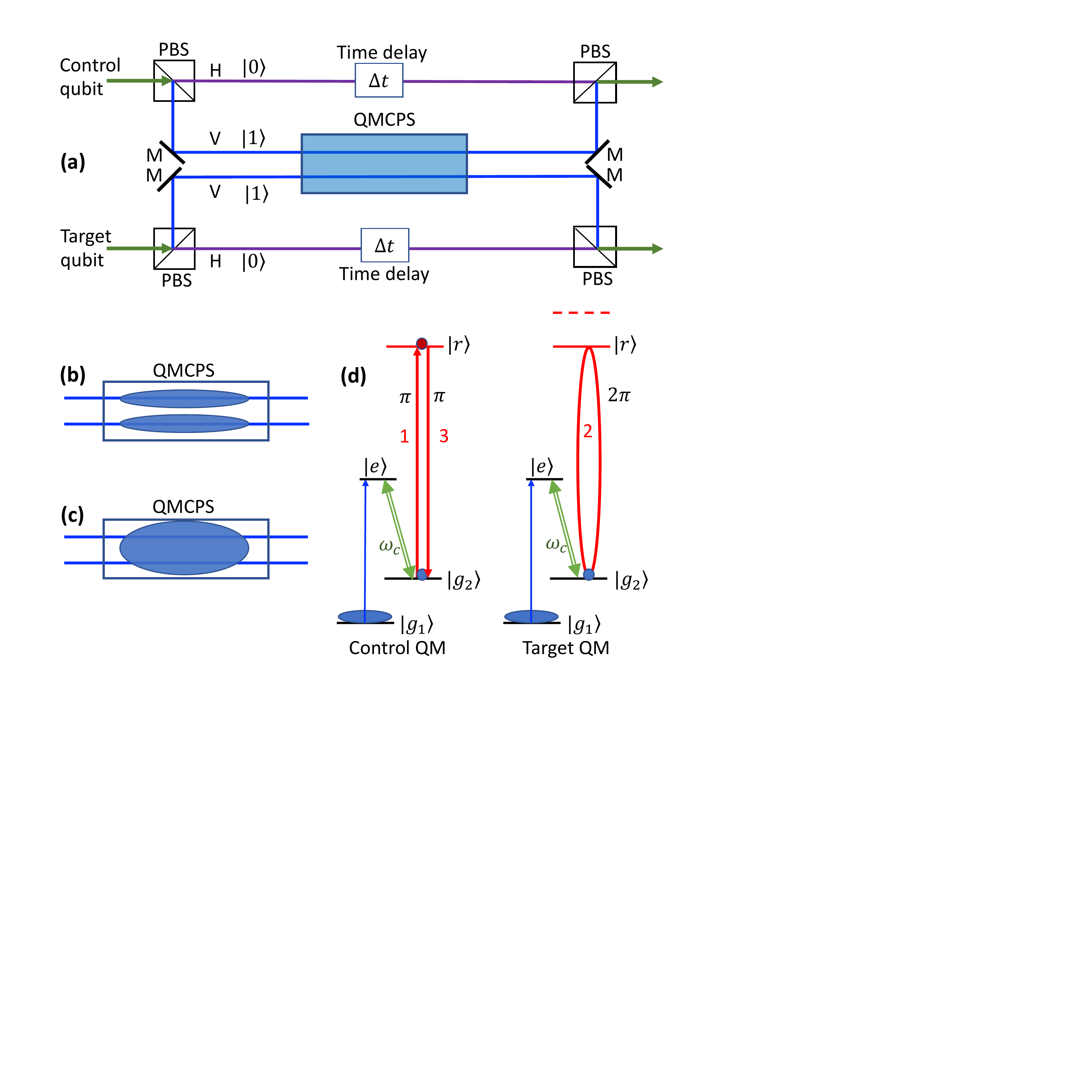}
\caption{CP gate implementation scheme 1. (a) The CP gate setup with polarization optics and quantum memory controlled phase shift (QMCPS) with two unoverlaping photon modes. PBS: polarizing beam splitter. M: mirror. (b) QMCPS implementation with two atomic ensembles. (c) QMCPS implementation with one atomic ensemble. (d) The atomic energy level diagram for QM and Rydberg blockade.}
\label{fig:CPgate1}
\end{figure}

\begin{figure}[t]
\includegraphics[width=1.0\linewidth]{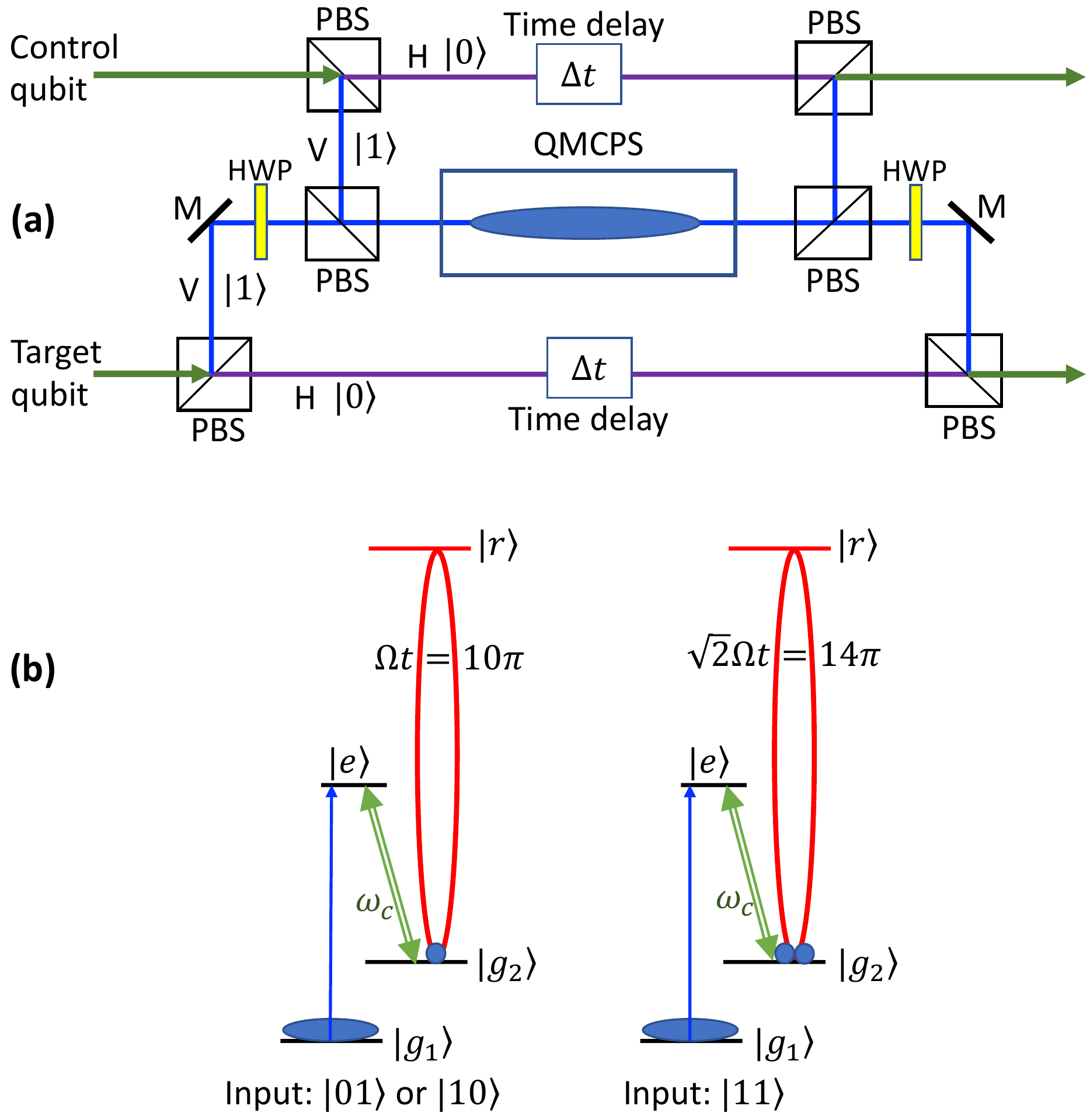}
\caption{CP gate implementation scheme 2. (a) The CP gate setup with polarization optics and QMCPS with only one shared atomic ensemble and two overlaping photon modes. HWP: half-wave plate. (b) The atomic energy level diagram for QM and Rydberg blockade operations.}
\label{fig:CPgate2}
\end{figure}

\emph{{\color{blue}CP gate scheme 1}.---}Figure \ref{fig:CPgate1} depicts our first scheme of the photon-atom QM mediated CP gate realization. The geometry is similar to the recently proposed implementation of atomic ensembles with non-blockade induced phase shift \cite{PhysRevA.91.030301}, but we here make use of Rydberg blockade effect. As shown in Fig.~\ref{fig:CPgate1}(a), we encode the single photon computational basis onto the two orthogonal polarizations: $|0\rangle=|H\rangle$ (horizontal) and $|1\rangle=|V\rangle$ (vertical). After passing through two polarizing beam splitters (PBSs), the polarizations of the control and target photons are spatially separated into four paths. The two V-polarized photon modes are injected into a QM controlled phase shift (QMCPS) unit. The QMCPS comprises two closely placed atomic ensembles with each for one photon mode as shown in Fig.~\ref{fig:CPgate1}(b), or one big atomic ensemble with two unoverlapping photon modes as shown in Fig.~\ref{fig:CPgate1}(c). After the QMCPS operation (explained later in detail), the stored photons are read out and combined with their H modes after another two PBSs. The time delays $\Delta t$ in the two H polarization paths are used to compensate the QMCPS operation time. When the input photon state is $|00\rangle$, where the first is the control qubit and second is the target qubit, both photons pass through the two H spatial paths without any interaction and the output is still $|00\rangle$. When the input states are $|01\rangle$, $|10\rangle$, and $|11\rangle$, the QMCPS operation is illustrated in Fig.~\ref{fig:CPgate1}(d). A QM \cite{Zhou:12, DuQM2019} with electromagnetically induced transparency (EIT) \cite{EIT-Harris, RevModPhys.77.633} usually involve three atomic states: two long-lived hyperfine ground states $|g_1\rangle$ and $|g_2\rangle$ and one excited state $|e\rangle$. A Rydberg state $|r\rangle$ with a large principle quantum number is used for Rydberg blockade \cite{NatPhys.5.110}. The qubit photons are on resonance at the transition $|g_1\rangle\leftrightarrow|e\rangle$. When the QM is in idle, all the atoms are prepared in the state $|g_1\rangle$ with presence of a control ($\omega_c$) laser beam on resonance to the transition $|g_2\rangle \leftrightarrow |e\rangle$. As a V-polarized qubit photon wave packet enters the QM, we switch off the control laser and convert the photonic state into the following entangled QM state \cite{RevModPhys.77.633}:
\begin{eqnarray}
|\mathrm{QM}\rangle=&\frac{1}{\sqrt{N_a}}[e^{i\phi_1}|g_2g_1g_1...g_1g_1\rangle+e^{i\phi_2}|g_1g_2g_1...g_1g_1\rangle \nonumber \\
&+...+e^{i\phi_{N_a}}|g_1g_1...g_1g_2\rangle],
\label{eq:QMstate}
\end{eqnarray}
where $N_a$ is the number of atoms. $\phi_j=\vec{k}\cdot \vec{r}_j$, with $\vec{k}$ the qubit photon wave vector, is the photon mode propagation phase at position $\vec{r}_j$ and stores the photon momentum information. After this QM writing operation on both memories, to attain the CP gate, in a similar manner to how CP gates are implemented in neutral atom quantum computing schemes \cite{NatPhys.5.110, PhysicsToday-Saffman, PhysRevLett.104.010503}, we apply the following three pulses sequentially: i) a $\pi$ Rydberg excitation pulse on resonance at the transition $|g_2\rangle \leftrightarrow |r\rangle$ to excite the control memory state $|\mathrm{QM}\rangle$ to the following collective Rydberg state:
\begin{eqnarray}
|\mathrm{QMR}\rangle=&\frac{1}{\sqrt{N_a}}[e^{i\phi_1}|rg_1g_1...g_1g_1\rangle+e^{i\phi_2}|g_1rg_1...g_1g_1\rangle \nonumber \\ 
&+...+e^{i\phi_{N_a}}|g_1g_1...g_1r\rangle];
\label{eq:QMR}
\end{eqnarray}
ii) a $2\pi$ pulse to the resonant transition $|g_2\rangle \leftrightarrow |r\rangle$ on the target memory, and iii) a second $\pi$ pulse to bring the control memory back to $|\mathrm{QM}\rangle$. After these three pulses, the control laser beams are switched back on to both memories and the QM state(s) is(are) then converted back to V-polarized photon(s). With the input state $|01\rangle$, all atoms in the control memory are in the state $|g_1\rangle$ without Rydberg excitation such that the target memory returns to its $|\mathrm{QM}\rangle$ with a negative sign after the $2\pi$ pulse. This negative sign is imprinted to the readout photon state, \textit{i.e.}, $|01\rangle\rightarrow -|01\rangle$. With the input $|10\rangle$, there is no excitation in the target memory and the control memory state obtains a negative sign after two $\pi$ pulses: $|10\rangle\rightarrow -|10\rangle$. In the case with the input $|11\rangle$, both memories are excited into the state $|\mathrm{QM}\rangle$. After the first $\pi$ pulse, the control memory is excited to its Rydberg state $|\mathrm{QMR}\rangle$, which induces a blue energy shift for the target memory Rydberg state $|r\rangle$ due to the dipole-dipole interaction and prevents Rydberg excitation in the target memory. This blockade effect makes the $2\pi$ pulse on the target memory unable to complete the excitation cycle, unable to gain a negative phase. After the second $\pi$ pulse, the control memory returns to its $|\mathrm{QM}\rangle$ with a negative sign. Overall, we obtain $|11\rangle\rightarrow -|11\rangle$ for the readout photons. In terms of the two-qubit basis $\{|00\rangle, |01\rangle, |10\rangle, |11\rangle \}$, the above CP gate can be described by a $4\times 4$ matrix
\begin{eqnarray}
\mathrm{CP}=\begin{bmatrix}
     1 & 0 & 0 & 0 \\
     0 & -1 & 0 & 0 \\
     0 & 0 & -1 & 0 \\
     0 & 0 & 0 & -1 
\end{bmatrix}.
\label{eq:CPgate}
\end{eqnarray}
%This scheme has similar geometry to the recently proposed implementation of two atomic ensembles with non-blockade induced phase shift \cite{PhysRevA.91.030301}. But our next scheme with only one atomic-ensemble QM described in the following section (Scheme 2) is significantly different.  

\begin{figure}[t]
\includegraphics[width=1.0\linewidth]{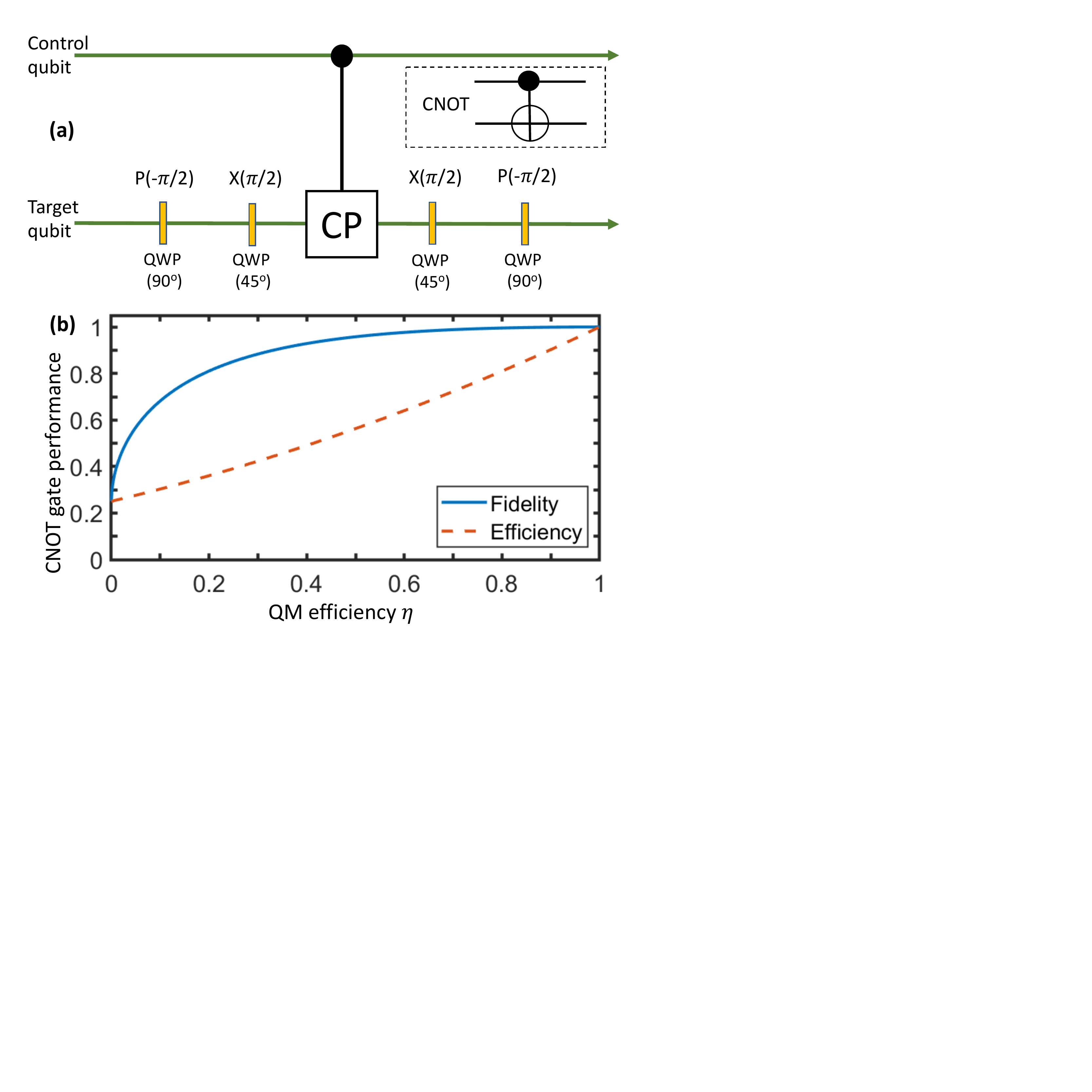}
\caption{CNOT gate realization. (a) Quantum circuit diagram of the CNOT gate: the CNOT gate is realized by sandwiching a CP gate in the middle of four target qubit operations. QWP: quarter-wave plate. The inset is the CNOT gate icon. (b) CNOT gate fidelity and efficiency as a function of QM efficiency $\eta$.}
\label{fig:CNOT}
\end{figure}

\emph{{\color{blue}CP gate scheme 2}.---}As an alternative scheme, Figure~\ref{fig:CPgate2} delineates our second CP gate architecture. Different from Scheme 1 with two closely placed QM atomic ensembles or two unoverlaping photon modes, here we have only one ensemble capable of storing two photonic modes which overlap in space. These two modes can be two orthogonal polarizations or two momentum modes \cite{DuQM2019}. For the purpose of illustration, in this scheme we focus on two polarization modes which have maximum spatial overlap in the single QM atomic ensemble, but keep in mind that the polarization modes can be converted into momentum modes as described in Ref. \cite{DuQM2019}. As outlined in Fig.~\ref{fig:CPgate2}(a), we transform the V polarization of the target photon into H polarization with a HWP and combine it with the V polarization of the control photon at a PBS (and reverse at output PBS). After the photon(s) is(are) stored inside the QM, we apply one single-atom $\Omega t=10\pi$ Rydberg excitation pulse, with $\Omega$ being the single-atom Rabi frequency and $t$ the pulse length. In the case with the input $|01\rangle$ or $|10\rangle$, only one atom is excited to $|g_2\rangle$ as shown in Fig.~\ref{fig:CPgate2}(b) and the overall QM state is described by Eq. (\ref{eq:QMstate}). Hence, the $10\pi$ Rydberg excitation pulse results in a negative sign to the QM state as well as to the retrieved photon. For the input $|11\rangle$ case, two atoms are excited to $|g_2\rangle$ and the QM state now becomes  
\begin{eqnarray}
|\mathrm{QM2}\rangle=\sqrt{\frac{2!(N_a-2)!}{N_a!}}[e^{i\phi_{12}}|g_2g_2g_1...g_1g_1\rangle+ \nonumber \\ 
e^{i\phi_{13}}|g_2g_1g_2...g_1g_1\rangle+...+e^{i\phi_{N_a-1,N_a}}|g_1g_1...g_1g_2g_2\rangle]
\label{eq:QMstate2}
\end{eqnarray}
with $\phi_{ij}=\phi_i-\phi_j$. For brief notation, we shorthand Eq.~(\ref{eq:QMstate2}) as $|\mathrm{QM2}\rangle=|g_2g_2\rangle$. With the same $10\pi$ Rydberg excitation pulse applied to two atoms, the blockade mechanism leads to an oscillation between $|g_2g_2\rangle$ and the symmetric Rydberg state $\frac{1}{\sqrt{2}}[|rg_2\rangle+|g_2r\rangle]$ with an effective Rabi frequency $\sqrt{2}\Omega$ \cite{NatPhys.5.115, NatPhys.8.790, PhysRevLett.128.123601}. Accordingly, the $\Omega t=10\pi$ Rydberg excitation pulse is effectively enhanced as $\sqrt{2}\Omega t=10\sqrt{2}\pi\simeq 14\pi$ by returning the QM state to $|\mathrm{QM2}\rangle$ with a $\pi$-phase shift. That is, $|11\rangle\rightarrow -|11\rangle$ for the readout photons. In this way, we obtain the same CP gate (Eq. (\ref{eq:CPgate})) as in Scheme 1. Yet, the technical advantages between Schemes 1 and 2 become apparent in the sense that the latter configuration not only improves Rydberg blockade effect due to the perfect spatial overlap of the two photonic modes, but requires only a single excitation pulse instead of three. We mark that, counting the difference between $10\sqrt{2}\pi$ and $14\pi$, the CP gate fidelity can still be more than 0.999.

\emph{{\color{blue}CNOT gate}.---}The CP gate can be transformed into a standard CNOT gate with additional target single-qubit operations, as represented by the quantum circuit of Fig.~\ref{fig:CNOT}(a). Here, $\mathrm{P}(-\pi/2)$ and $\mathrm{X}(\pi/2)$ are given by
\begin{eqnarray}
\mathrm{P}(-\frac{\pi}{2})=\begin{bmatrix}
     1 & 0 & 0 & 0 \\
     0 & -i & 0 & 0 \\
     0 & 0 & 1 & 0 \\
     0 & 0 & 0 & -i 
\end{bmatrix},\;
%\label{eq:Pgate}
%\end{eqnarray}
%$\mathrm{X}(\pi/2)$ is a $\pi/2$ rotation gate
%\begin{eqnarray}
\mathrm{X}(\frac{\pi}{2})=\begin{bmatrix}
     1 & -i & 0 & 0 \\
     -i & 1 & 0 & 0 \\
     0 & 0 & 1 & -i \\
     0 & 0 & -i & 1 
\end{bmatrix}.
\label{eq:Xgate}
\end{eqnarray}
For a photonic polarization qubit, an arbitrary unitary transformation can be realized with a combination of HWPs and quarter-wave plates (QWPs) by properly aligning their slow-fast axes \cite{Barz_2015, Saleh1991}. The single-qubit phase gate $\mathrm{P}(-\pi/2)$ is realized by a QWP whose fast axis is aligned along the V-polarization direction. The $\mathrm{X}(\pi/2)$ rotation gate is achieved by a QWP whose fast axis is aligned at 45$^o$ with respect to the H- polarization direction. Following the quantum circuit, we get   
\begin{eqnarray}
\mathrm{CNOT}=\begin{bmatrix}
     1 & 0 & 0 & 0 \\
     0 & 1 & 0 & 0 \\
     0 & 0 & 0 & 1 \\
     0 & 0 & 1 & 0 
\end{bmatrix}.
\label{eq:CNOT}
\end{eqnarray}

The QM efficiency, which is always less than unity in reality and can be modeled as photonic loss, plays an important role on the CNOT gate performance. In Fig.~\ref{fig:CNOT}(b), we plot the CNOT gate fidelity and efficiency as a function of QM efficiency $\eta$. While the fidelity remains as high as $>$0.9 as the QM efficiency $\eta$ drops to 0.33, the gate efficiency decreases to 0.44. This marks a significant difference between photonic and other quantum computing platforms. For the trapped-ion and atom-array systems, their gate fidelities depend strongly on the control noise as it reduces a pure qubit state into a mixed one. In our photon-atom hybrid system, the coupling between the qubit Hilbert space and environment is only caused by the loss, and the lost photons disappear into the environment but are not detected by single-photon counters. As a result, the QM loss does not affect the fidelity much, but reduces the state generation efficiency as a cost. %For $\eta > 0.4$, the gate fidelity is not sensitive to the QM efficiency because of the postselection renormalization. 

\begin{figure}[t]
\includegraphics[width=0.85\linewidth]{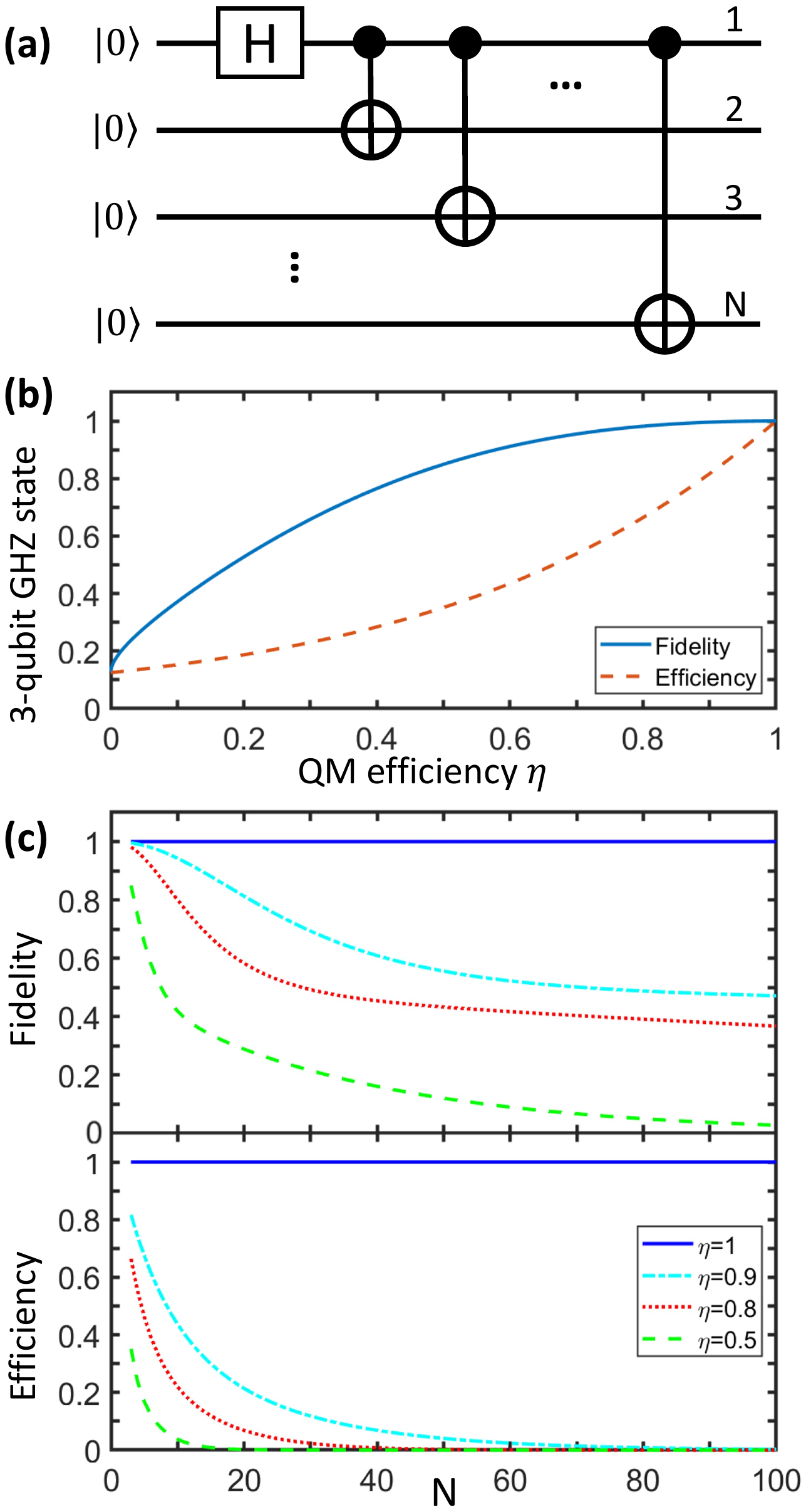}
\caption{GHZ state generation. (a) The quantum circuit for generating N-qubit GHZ state. (b) The 3-qubit GHZ state fidelity and generation efficiency as a function of QM efficiency $\eta$. (c) N-qubit GHZ state fidelity and generation efficiency as a function of N.}
\label{fig:GHZ}
\end{figure}

\emph{{\color{blue}Greenberger–Horne–Zeilinger (GHZ) state generation}.---} As an example of application, we apply the QM-mediated CNOT gates and linear optics to generate an N-photon GHZ state \cite{arxiv.0712.0921, PhysRevLett.82.1345}, $\frac{1}{\sqrt{2}}[|000...\rangle+|111...\rangle]$. Figure~\ref{fig:GHZ}(a) is the quantum circuit with the initial (input) unentangled state prepared as $|000...\rangle$, involving N-1 CNOT gates. The Hadamard (H) gate transforms the first qubit from $|0\rangle$ to $\frac{1}{\sqrt{2}}[|0\rangle+|1\rangle]$, and can be implemented by a HWP with its fast axis aligned at $22.5^o$ to the H-polarization axis. The fidelity and efficiency of yielding a three-photon GHZ state as a function of single QM efficiency $\eta$ are given in Fig.~\ref{fig:GHZ}(b). To have the fidelity $>$0.9, it requires $\eta>0.58$ where the state generation efficiency is 0.42. To investigate the scalability, we plot the state fidelity and generation efficiency as a function of N for different $\eta$ in Figs.~\ref{fig:GHZ}(c) and (d), respectively. As one can see, when N=100 a fidelity $>$0.47 is still achievable for $\eta=0.9$, but the generation efficiency reduces sharply as N increases.

\emph{{\color{blue}Discussion}.---}In the above description, the quantum circuit elements are spatially distributed and connected via optical modes. Thus, our hybrid photon-atom scheme can be used for distributed quantum computing. While the current researches in quantum computing and networks are nearly isolated and there is a lack of protocol for networking distributed multiple-qubit quantum computers, our scheme provides a natural quantum network interface between flying photons and local atomic quantum nodes. It may be possible to construct a distributed quantum computer with this photon-atom hybrid architecture, or cloud quantum computing with various quantum computers.

\begin{figure}[t]
\includegraphics[width=0.85\linewidth]{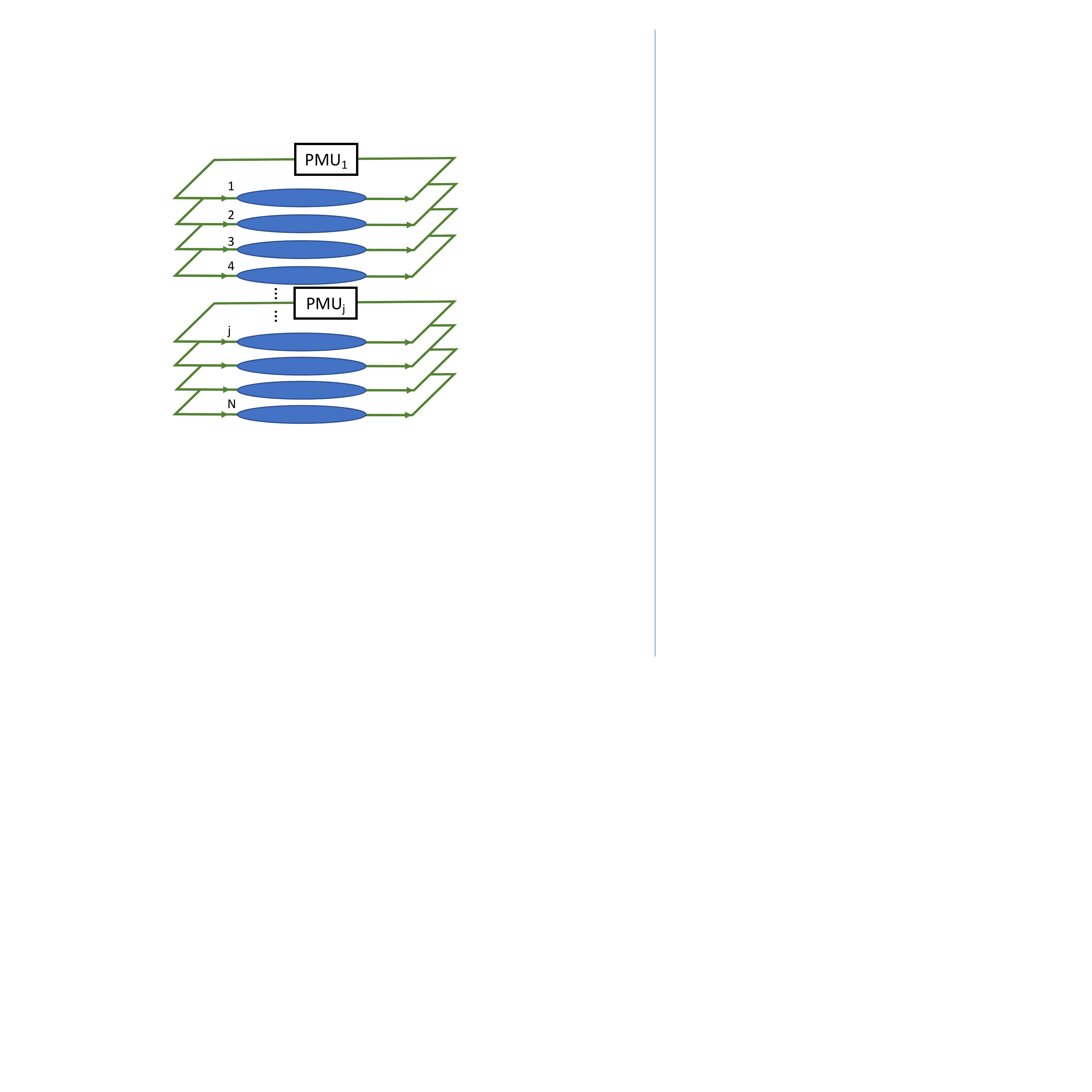}
\caption{Schematics of time-line quantum computing with an array of N atomic ensemble QMs. At each QM, the readout photon is fed back to the input port after passing through a polarization manipulation unit (PMU). Each QM can store both polarization modes.}
\label{fig:QC-Timeline}
\end{figure}

It is worthy to point out that our scheme can also be used to construct a time-line quantum computer by recycling the QMs, similarly to those platforms with trapped ions and neutral atoms. Figure \ref{fig:QC-Timeline} schematics such an N-qubit quantum computer structure with a one-dimensional (or two-dimensional) array of N QM atomic ensembles. For each QM, its readout photon is sent back to the QM after a programmable unitary transformation -- a polarization manipulation unit (PMU), which can be realized by a combination of HWPs, QWPs and other linear optics. The nonlinear controlled gate interaction between any two qubit memories can be mediated by the Rydberg blockade effect. The time-line programmable depth, or the effective coherence time, of such a quantum computer is limited by the QM efficiency. The number of programmable steps is proportional to $-1/\ln{\eta}$. 

As shown above, QMs are essential for the proposed schemes, providing conversion interfaces between single-photon polarization qubits and atomic states. It is extremely challenging to implement efficient QM with single atom or ion, while an atomic ensemble has a collective enhancement under the phase-matching condition. Among various schemes including photon echo \cite{Nature.465.1052, Cho:16} and off-resonance Raman interaction \cite{NaturePhotonics.4.218, PhysRevLett.107.053603}, so far EIT ground-state QM with laser-cooled atoms \cite{Du2DMOT2012} has demonstrated the highest efficiency ($\eta>$85\%) for single photon polarization qubits with a fidelity of more than 99\% \cite{DuQM2019}. For a single polarization channel, the memory efficiency can be as high as 90.6\% \cite{DuQM2019}. One technical challenge to implement the CP gate scheme 1 and time-line quantum computing is to arrange multiple  cigar-shaped cold atomic ensembles closely enough for Rydberg blockade effect, or arrange multiple closely spacing but unoverlaping photon modes in a big atomic ensemble with each having sufficient optical depth. For heavy alkali atoms widely used for laser cooling and trapping, such as Rb and Cs, their Rydberg interaction distance can be $>$40 $\mu$m for a large principle quantum number ($n\geq 200$) \cite{RevModPhys.82.2313}. Atom chip technique \cite{PhysRevA.70.053606, Folman2002} maybe a solution to prepare array of atomic ensembles. Recently, trapping hundreds of microscopic atomic ensembles in optical tweezer arrays has been demonstrated \cite{npjqi.6.54}, which could be used for atomic-ensemble based Rydberg qubits \cite{PhysRevLett.128.123601, PhysRevLett.127.050501}. In the CP gate scheme 2, the control and target photonic modes overlap inside QM, leading to more efficient Rydberg blockade effect. Different from the platforms with trapped single ions and neutral atoms whose coherence limits the computation duration, in our scheme the computation time is only limited by the photon loss, but not by the QM lifetime as it can be recycled and only requires time to complete a CP operation. 

Our proposed solution incorporates the already established photonic linear manipulation and neutral atom nonliner Rydberg interaction, encompassing building blocks not only for quantum computers, but also extending its capability to quantum networks. An attractive feature of this idea is that it can be spatially and temporally distributed. We acknowledge that our initial model has unresolved physics such as QM efficiency, limited Rydberg blockade radius, and imperfect pulses which may degrade two qubit fidelity. Nevertheless, our scheme offers scalability for both single-qubit and two-qubit controlled gates.

\begin{acknowledgments}
\textbf{Acknowledgments.---}
%E.O. would like to thank LPS for support. 
X.L. and J.W. acknowledge support from DOE (DE-SC0022069). S.D. acknowledges support from AFOSR (FA9550-22-1-0043) and NSF (CNS-2114076).
\end{acknowledgments}

\bibliography{QC}

\end{document}